\documentclass[aps,prb,twocolumn,superscriptaddress,showpacs]{revtex4}
\usepackage[pdftex]{graphicx}
\usepackage{amsmath}
\usepackage{amssymb}

\begin{document}

\title{Transport in gapped bilayer graphene: the role of potential fluctuations}

\author{K. Zou}
\affiliation{Department of Physics, The Pennsylvania State
University, University Park, PA 16802}
\author{J. Zhu}
\affiliation{Department of Physics, The Pennsylvania State University, University Park, PA 16802}



\begin{abstract}
We employ a dual-gated geometry to control the band gap $\Delta$ in bilayer graphene and study the temperature dependence of the resistance at the charge neutrality point, $R_{\mathrm{NP}}(T)$, from 220 to 1.5 K. Above 5 K, $R_{\mathrm{NP}}(T)$ is dominated by two thermally activated processes in different temperature regimes and exhibits exp($T_{3}/T)^{1/3}$ below 5 K. We develop a simple model to account for the experimental observations, which highlights the crucial role of localized states produced by potential fluctuations. The high temperature conduction is attributed to thermal activation to the mobility edge. The activation energy approaches $\Delta/2$ at large band gap. At intermediate and low temperatures, the dominant conduction mechanisms are nearest neighbor hopping and variable-range hopping through localized states. Our systematic study provides a coherent understanding of transport in gapped bilayer graphene.
\end{abstract}

\maketitle

Bilayer graphene is a unique two-dimensional (2D) material with a tunable bandgap. A perpendicular electric field breaks the inversion symmetry between the two graphene layers and results in a field-dependent bandgap~\cite{McCann2006gap,Min2007,Gava2009}. Its experimental signatures have been observed by infrared spectroscopy~\cite{Zhang2009,Mak2009,Kuzmenko2009,ZhangLM2008} and angle-resolved photoemission~\cite{Ohta2006}, but remain incomplete and perplexing in transport~\cite{Castro2007,Oostinga2008,Xia2009,Kim2009}. Near room temperature, Xia et al.  observes thermally activated conduction and attributes it to Schottky barriers at the electrode-gapped bilayer interface~\cite{Xia2009}. In the mK regime, Oostinga et al. reports variable-range hopping~\cite{Oostinga2008}. To date, systematic investigations combining high and low temperatures are lacking and a coherent understanding of conduction in gapped bilayer has yet to emerge.

In this work, we control the band gap in bilayer graphene using top and bottom gates, and measure the temperature-dependent resistance at the charge neutrality point (CNP) $R_{\mathrm{NP}}(T)$ as a function of the gap, in the temperature range of 1.5 K $< T < $ 220 K. We develop a model to explain the data, which highlights the essential role of localized states produced by potential fluctuations. Our data point to three conduction mechanisms: thermal activation to the mobility edge at high temperatures, nearest neighbor hopping at intermediate temperatures and variable-range hopping at low temperatures.

We fabricate SiO$_{2}$/HfO$_{2}$ dual-gated bilayer graphene field effect transistors using procedures previously described in Ref.~\onlinecite{Zou2009}. 30 nm HfO$_{2}$ is deposited on single or bilayer graphene by atomic layer deposition and used as the topgate dielectrics. We have achieved high mobility $\mu$ of 9,000 -- 16,000 cm$^{2}$/Vs on single layer graphene~\cite{Zou2009}. Here, on dual-gated bilayer, $\mu$ ranges from 1,500 to 6,000 cm$^{2}$/Vs, which is generally lower than $\mu$ up to 12,000 cm$^{2}$/Vs observed in our pristine bilayer samples.  Raman spectra on dual-gated bilayer devices show no visible D band, indicating minimal defect creation (Fig. S4 in Ref.~\onlinecite{support}). The gating efficiency of the topgate is approximately 2.8$\times$10$^{12}$/cm$^{2}$ per volt from Hall measurements, which is $\sim$ 40 times of the efficiency of the 290 nm SiO$_{2}$ backgate. The gating range of the topgate is greater than 1.4$\times$10$^{13}$/cm$^{2}$.

\begin{figure}
\includegraphics[angle=0,width=2.2in]{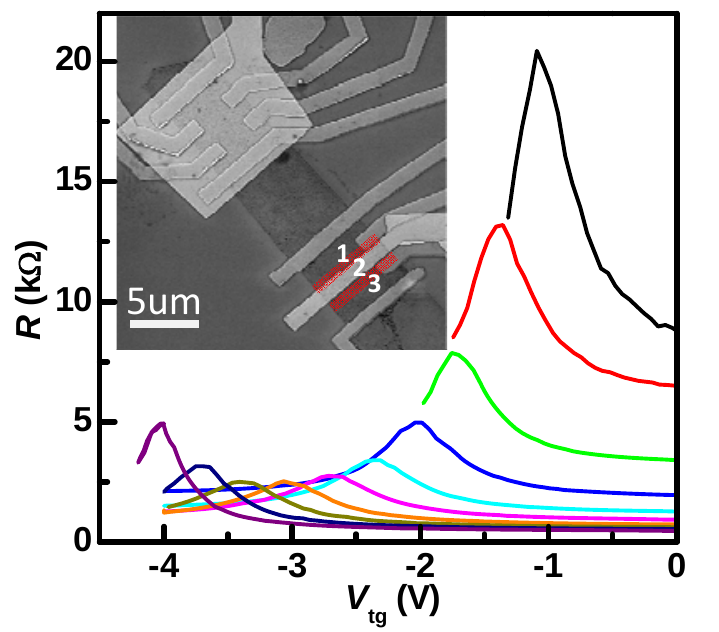}
\vspace{-0.1in}
 \caption[]{(Color online) $R(V_{\mathrm{tg}})$ at fixed $V_{\mathrm{bg}}$ of sample B1 at 10 K. From left to right, $V_{\mathrm{bg}}$= +30 V to -60 V at 10 V steps. Inset: a SEM picture of a device. The mobility $\mu$ is approximately 1,500 cm$^{2}$/Vs for sample A1, A2 and 3,000 cm$^{2}$/Vs for sample B1, B2.
\label{device}}
\vspace{-0.2in}
\end{figure}

A scanning electron microscope (SEM) micrograph of a dual-gated device is shown as an inset to Fig.~\ref{device}. Two types of topgates are used. In samples A1 and A2, the area between the two voltage probes is completely covered by the topgate electrode (top of inset), so that it does not include the interfacial resistances between the gapped and ungapped area. In samples B1 and B2, the coverage is partial (bottom of inset). All four measurements yield similar behavior, indicating that the resistance of gapped bilayer is dominated by the bulk.

To obtain the $T$-dependence of the charge neutrality point resistance $R_{\mathrm{NP}}(T)$, we fix the backgate voltage $V_{\mathrm{bg}}$ and sweep the topgate $V_{\mathrm{tg}}$. The sweep is repeated at different $V_{\mathrm{bg}}$'s and temperatures. Figure~\ref{device} shows $R(V_{\mathrm{tg}})$ of sample B1 at 10 K. From left to right, $V_{\mathrm{bg}}$ changes from +30 V to -60 V at 10 V steps. The maximum resistance of each curve corresponds to the CNP for that specific pair of ($V_{\mathrm{bg}}$, $V_{\mathrm{tg}}$) settings. The minimum of all $R_{\mathrm{NP}}$ is found at roughly $V_{\mathrm{bg0}}$ = 0 V and $V_{\mathrm{tg0}}$ = 3.1 V in Fig.~\ref{device}, which corresponds to the condition where the average band gap $\Delta$ = 0~\cite{Zhang2009}. The offsets $V_{\mathrm{tg0}}$ and $V_{\mathrm{bg0}}$ comes from unintentional chemical doping. At the CNP, $\Delta$ was shown to increase with the electric displacement field $D = D_{\mathrm{bg}} = -D_{\mathrm{tg}} = \epsilon_{\mathrm{bg}}V^{\prime}_{\mathrm{bg}}/d_{\mathrm{bg}}$, where $\epsilon_{\mathrm{bg}}$ and $d_{\mathrm{bg}}$ are the dielectric constant and thickness of the backgate oxide respectively, and $V^{\prime}_{\mathrm{bg}} = V_{\mathrm{bg}}-V_{\mathrm{bg0}}$ is the effective backgate voltage~\cite{McCann2006gap,Zhang2009}. Using  $\epsilon_{\mathrm{SiO2}}$ = 3.9 and $d_{\mathrm{SiO2}}$ = 290 nm, we obtain $D$ = 1.34 V/nm for $V^{\prime}_{\mathrm{bg}}$ = 100 V. In samples B1 and B2, the resistance of the dual-gated area 2 is calculated by subtracting the resistance of areas 1 and 3 (Fig.~\ref{device} inset), which is obtained by sweeping $V_{\mathrm{bg}}$ while grounding $V_{\mathrm{tg}}$ at each temperature. Standard lock-in techniques are used with excitation current 0.2 nA to 50 nA. Low currents are carefully chosen for high-resistance measurements to avoid current heating.

\begin{figure}
\includegraphics[angle=0,width=2.4in]{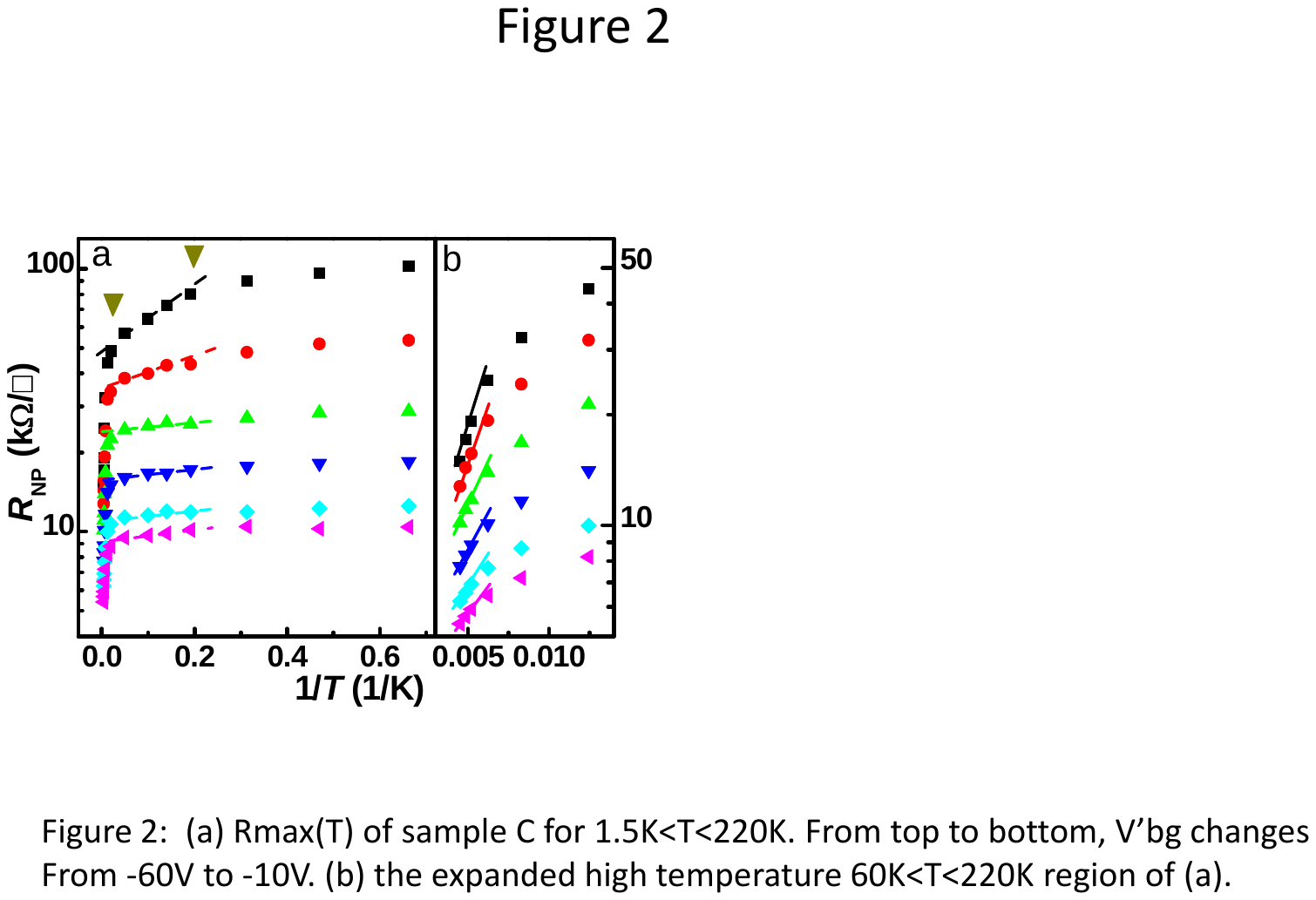}
\vspace{-0.1in}
 \caption[]{(Color online) (a) $R_{\mathrm{NP}}(T)$ of sample B1 in a semi-log vs. 1/$T$ plot in the temperature range of 1.5 K $< T < $ 220 K. From top to bottom, $V^{\prime}_{\mathrm{bg}}$ = -60 V to -10 V at 10 V steps. The triangles mark 50 K and 5 K respectively. (b) The expanded region of 70 K$< T <$ 220 K in (a). The solid and dashed lines in (a) and (b) are guide to the eye.
\label{RNPdata}}
\vspace{-0.2in}
\end{figure}

$R_{\mathrm{NP}}(T)$ of sample B1 is shown in a semi-log vs. 1/$T$ plot in Fig.~\ref{RNPdata}(a). From top to bottom, $V^{\prime}_{\mathrm{bg}}$ changes from -60 V to -10 V at 10 V steps. $R_{\mathrm{NP}}(T)$ shows an insulating $T$-dependence in the whole temperature range. The high-$T$ region is expanded in Fig.~\ref{RNPdata}(b), where log$R_{\mathrm{NP}}$ vs $1/T$ shows a linear trend (solid lines) at $T >$ 150 K.  This trend suggests thermally actived conduction. From 150 K to 50 K, the slopes continue to decrease, and become approximately constant again between 50 K and 5 K albeit with much smaller values (dashed lines in Fig.~\ref{RNPdata}(a)). This suggests another thermally activated process with a much smaller activation energy. Below 5 K, $R_{\mathrm{NP}}(T)$ remains nearly $T$-independent for small $|V^{\prime}_{\mathrm{bg}}|$ and increases slowly with decreasing $T$ for larger  $|V^{\prime}_{\mathrm{bg}}|$. Similar behavior is observed on all four samples.

\begin{figure}
\includegraphics[angle=0,width=2.2in]{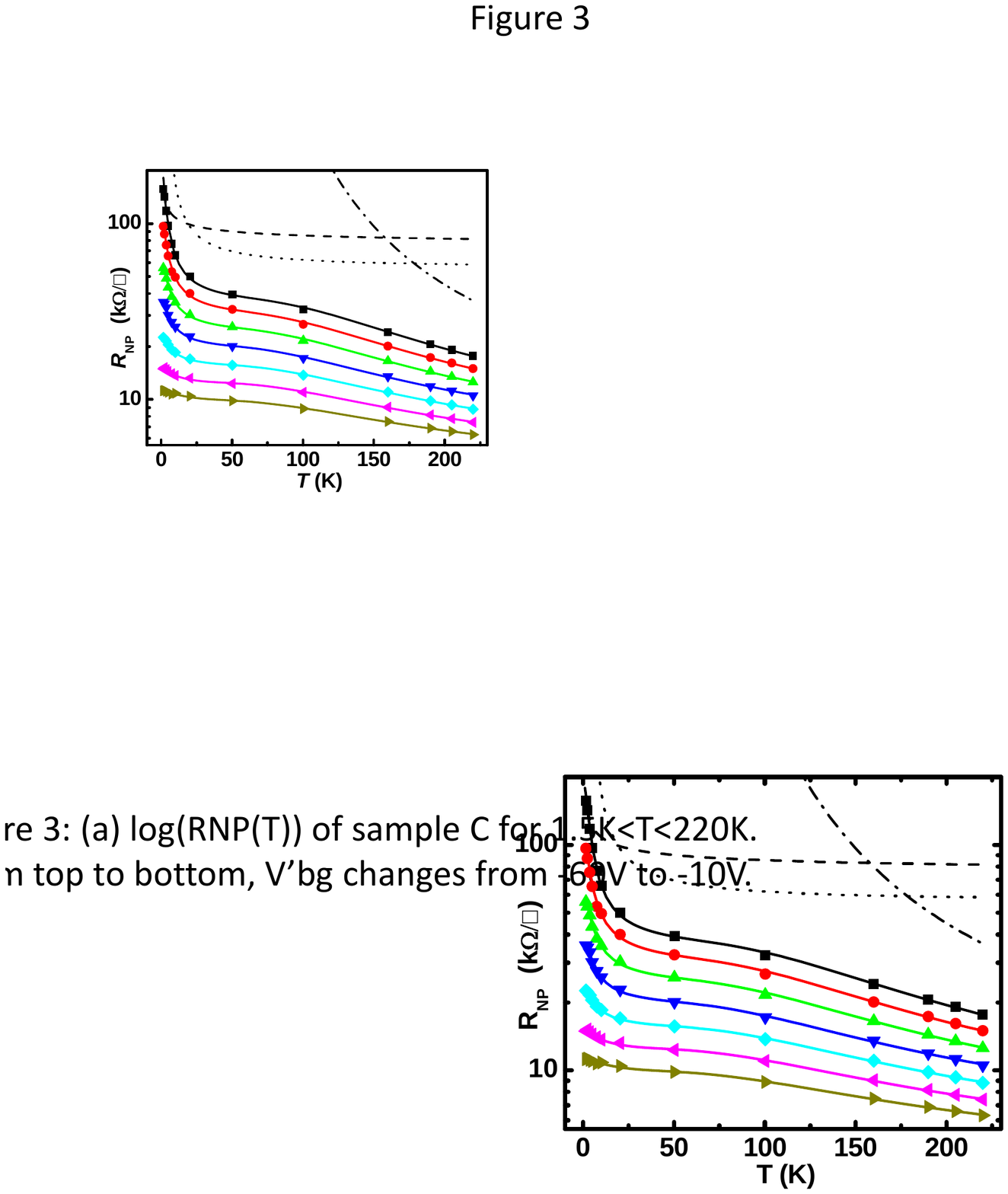}
\vspace{-0.1in}
 \caption[]{(Color online) $R_{\mathrm{NP}}(T)$ of sample A1 in the temperature range of 1.5 K $< T < $ 220 K. From top to bottom, $V^{\prime}_{\mathrm{bg}}$ = -90 V to -30 V at 10 V steps. The solid lines are fittings to Eq.~\ref{fit}. The 3 terms in Eq.~\ref{fit} are plotted as dash-dotted, dotted, and dashed lines respectively for the top curve to show the contribution of each term at different temperatures.
\label{RNPfit}}
\vspace{-0.2in}
\end{figure}

These observations are reminiscent of impurity conduction in doped semiconductors~\cite{Shklovskii1984,Mott1993}, where localized states produced by impurities reside in the band tail and are separated from delocalized states by a mobility edge $E_{\mathrm{c}}$. In such systems, thermal activation to the mobility edge dominates the high-temperature conduction. As $T$ decreases, electrons hop through localized states, overcoming the energy difference between adjacent neighbors, i. e., nearest neighbor hopping. At even lower $T$, electrons hop through localized states far from each other via the assistance of phonons of energy $k_{\mathrm{B}}T$, giving rise to variable-range hopping~\cite{Shklovskii1984,Mott1993}.

Localized states due to disorder also exist in bilayer graphene~\cite{Nilsson2007}. Scanning tunneling microscope (STM) measurements show the potential of the CNP in bilayer can fluctuate between $\pm$ 40 meV~\cite{Deshpande2009}. Similar observations have been made on single-layers as well~\cite{Martin2008,Zhang2009STM}. The resulting electron and hole puddles are typically a few tens of nm in lateral size~\cite{Deshpande2009,Martin2008,Zhang2009STM}. In a gapped bilayer, these puddles are confined and separated from the mobility edge above the potential fluctuations by a binding energy. In a zero-gap bilayer, Klein tunneling may provide a parallel conduction path to thermally activated processes~\cite{Katsnelson2006,Rossi2010}.

The $T$-dependence of $R_{\mathrm{NP}}(T)$ in Fig.~\ref{RNPdata} and the above reasoning led us to propose a fitting including the following three terms:

\begin{eqnarray}
\notag R_{\mathrm{NP}}(T)^{-1}=R_{1}^{-1}exp[-E_{1}/k_{\mathrm{B}}T]
\\+R_{2}^{-1}exp[-E_{2}/k_{\mathrm{B}}T]
\notag \\+R_{3}^{-1}exp[-(T_{3}/T)^{1/3}]
\label{fit}
\end{eqnarray}
where $E_{1}$, $E_{2}$ and $T_{3}$ represent the two activation and the hopping energy respectively and $R_{1}$, $R_{2}$ and $R_{3}$ are the corresponding resistance coefficients.  Eq.~\ref{fit} produces an excellent description of $R_{\mathrm{NP}}(T)$ in the whole temperature range in all samples. Each parameter plays a different role in the fitting~\cite{R1}. The fitting curves of sample A1 are shown in Fig.~\ref{RNPfit}. The combination of different energy scales accurately captures the shoulder seen around 50 K (Fig.~\ref{RNPfit}), and the bend around 5 K (Fig.~\ref{RNPdata}(a)) in all samples.

\begin{figure}
\includegraphics[angle=0,width=2in]{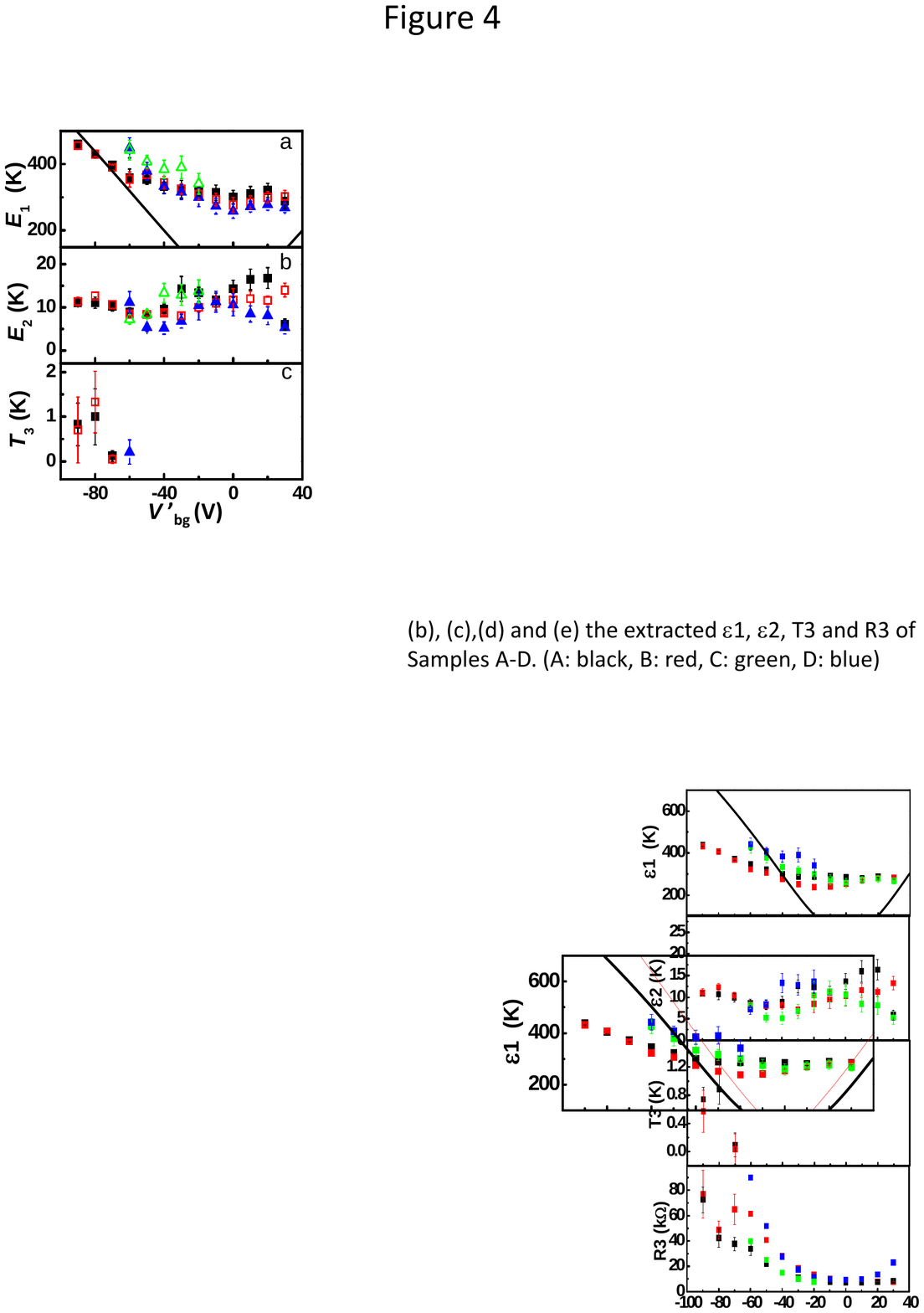}
\vspace{-0.1in}
 \caption[]{(Color online) (a)(b)(c) $E_{\mathrm{1}}$, $E_{\mathrm{2}}$, and $T_{\mathrm{3}}$ vs. $V^{\prime}_{\mathrm{bg}}$ for sample A1 (solid black square), A2 (hollow red square), B1 (solid blue triangle) and B2 (hollow green triangle). The theoretical $\Delta/2$~\cite{McCann2006gap} is plotted as a solid line in (a) for comparison.
\label{result}}
\vspace{-0.2in}
\end{figure}

The extracted $E_{1}$ of all four samples are plotted in Fig.~\ref{result}(a) and show similar trends. $E_{1}$ remains approximately constant at 250 K for small $|V^{\prime}_{\mathrm{bg}}|$ (or equivalently small $D$), and starts to increase at larger $|V^{\prime}_{\mathrm{bg}}|$, reaching $\sim$ 450 K at $V^{\prime}_{\mathrm{bg}}$ = -90 V in samples A1 and A2. Also plotted in Fig.~\ref{result}(a) is the calculated $\Delta$/2 of Ref.~\onlinecite{McCann2006gap} (solid black line). At small $|V^{\prime}_{\mathrm{bg}}|$, $E_{1}$ differs from the small $\Delta$/2 predicted by theory but approaches theory with increasing $|V^{\prime}_{\mathrm{bg}}|$. In particular, a finite $E_{1}$ at $V^{\prime}_{\mathrm{bg}}$ = 0 V, where $\Delta$ = 0, suggests a different activation mechanism. We attribute $E_{1}$ to the activation barrier to the mobility edge, i. e., $E_{1}=E_{\mathrm{c}}-E_{\mathrm{F}}$. Since at $\Delta$ = 0, the Fermi level $E_{\mathrm{F}}$ lies approximately in the middle of the fluctuating disorder potential $\Phi$, we associate $E_{1}$ = $E_{\mathrm{c}}$ with the root mean square (RMS) amplitude of the potential fluctuation, i. e., $\Phi_{\mathrm{RMS}}$ = $E_{1}$ = 250 K or 21.5 meV. Independently, $\Phi_{\mathrm{RMS}}$ in the STM data of Ref.~\onlinecite{Deshpande2009} was found to be 19 meV~\cite{support}, in good agreement with results obtained here. In this scenario, all carriers residing in the puddles will be excited above $E_{\mathrm{c}}$ at very high temperatures. Estimating the carrier density in the puddles to be $n = \Phi_{\mathrm{RMS}}(2m^{*}/\pi\hbar^{2})$ = 6$\times$$10^{11}/\mathrm{cm^{2}}  (m^{*} = 0.033m_{\mathrm{e}})$ and using $\mu$ = 1,500 -- 3,000 cm$^{2}$/Vs for all samples, we expect $R_{1} = 1/ne\mu$ to be 3500 -- 7000 $\Omega/\square$. $R_{1}$ extracted from the fittings ranges 3,000 -- 5,000 $\Omega/\square$, in excellent agreement with our estimates~\cite{R1}.

The above analysis should also hold for a small band gap $\Delta < \Phi_{\mathrm{RMS}}$, which explains the nearly flat $E_{1}$ at small $|V^{\prime}_{\mathrm{bg}}|$ in Fig.~\ref{result}(a). As $\Delta$ further increases, we expect $E_{1}$ to be gradually dominated by the band gap itself and eventually approach $\Delta$/2. Our results in Fig.~\ref{result}(a) support this trend. Small differences likely arise from the  details of the disorder potential $\Phi$ in individual samples. The above model also explains the insulating $T$-dependence widely observed near the CNP in backgated bilayers~\cite{Morozov2008,Feldman2009,Zou,Hwang2010}.

\begin{figure}
\includegraphics[angle=0,width=2.3in]{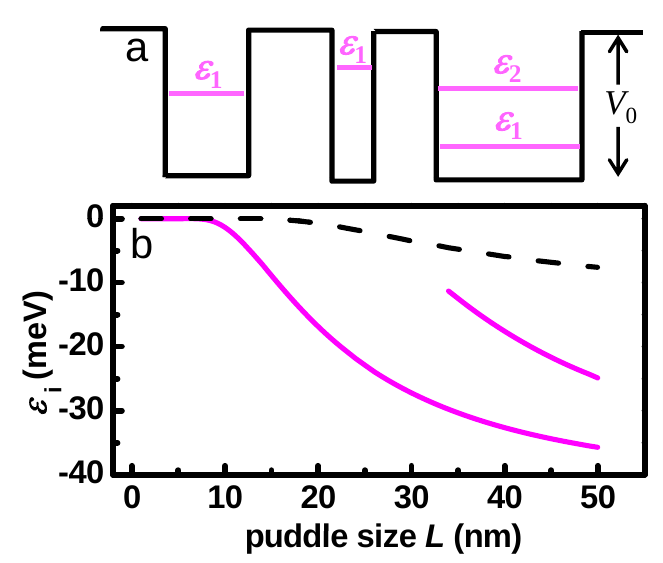}
\vspace{-0.1in}
 \caption[]{(Color online) (a) A simple model of the potential fluctuations in bilayer. The bound state of each cylindrical well is schematically shown. A wide well may have more than one bound state. (b)The bound state energy $\varepsilon_{\mathrm{i}}(L)$ for $V_{0}$ = 43 meV (solid magenta line) and $V_{0}$ = 13 meV (dashed black line).
\label{band}}
\vspace{-0.2in}
\end{figure}

As $T$ decreases, the above thermally activated conduction becomes inefficient; our data in Fig.~\ref{RNPdata} show that other low-energy processes gradually take over from 150 K to 50 K. Between 50 K $> T >$ 5 K, the rise of $R_{\mathrm{NP}}$ ($T$) is largely given by the 2$^{nd}$ exponental term in Eq.~\ref{fit} (Fig.~\ref{RNPfit}). The extracted activation energy $E_{2}$ is shown in Fig.~\ref{result}(b). $E_{2}$ is nearly $V^{\prime}_{\mathrm{bg}}$-independent and averages $\sim$ 10 K, or 25-50 times smaller than $E_{1}$, in all samples. In doped semiconductors, $E_{2}$ is associated with the hopping conduction between nearest neighboring impurity states. Percolation theory gives  $E_{2}$ = $<\varepsilon_{\mathrm{ij}}>$ = $<\varepsilon_{\mathrm{i}}$-$\varepsilon_{\mathrm{j}}>$ is the average energy difference between neighboring localized states on the percolation path~\cite{Shklovskii1984}. Similar to the localized states in doped semiconductors, in a gapped bilayer, electron and hole states near the band edges are quantized and localized due to small puddle size~\cite{E2note}. These localized states may support a hopping conduction that is more effecitive than the thermal activation above the mobility edge at $k_{\mathrm{B}}T << E_{1}$. In this scenario, the hopping is expected to occur between nearest neighbors (NNH) when the temperature is not too low, and transitions to variable-range hopping (VRH) at the lowest temperatures~\cite{Shklovskii1984}. These two hopping processes can account for the exp$[-E_{2}/k_{\mathrm{B}}T]$ and the exp$[-(T_{3}/T)^{1/3}]$ terms necessary to describe our data.

The following analysis provides an order of magnitude estimate of $E_{2}$ = $<\varepsilon_{\mathrm{ij}}>$ in our samples. We assume the distribution of the puddle size $L$ in our samples is similar to that in Ref.~\onlinecite{Deshpande2009}, since both samples are on SiO$_{2}$ substrates and have similar $\Phi_{\mathrm{RMS}}$. This distribution is estimated by analyzing the potential map of the CNP in Ref.~\onlinecite{Deshpande2009}. A histogram of $L$ is generated and modeled by a Gaussian distribution as shown in Fig. S2(b)~\cite{support}. The best fit yields $L$ = (13$\pm$8.4) nm, consistent with other STM reports~\cite{Zhang2009STM}. In our simplified model, the disorder potential $\Phi$ is approximated by a random network of cylindrical wells with a single depth $V_{0}$ but varying diameter $L$ as shown in Fig.~\ref{band}(a), neglecting the shape variation and the distribution of $V_0$. The bound state energy $\varepsilon_{\mathrm{i}}$ of each well is calculated using a constant effective mass $m^{*} = 0.033m_{\mathrm{e}}$~\cite{Mass}. Fig.~\ref{band}(b) plots $\varepsilon_{\mathrm{i}}$ vs. $L$ for two representative $V_{0}$'s. The solid curve corresponds to $V_{0}$ = 43 meV, which is 2$\Phi_{\mathrm{RMS}}$ in our samples. The dashed curve corresponds to $V_{0}$ = 13 meV. We obtain the average $<\varepsilon_{\mathrm{ij}}>$ using $\varepsilon_{\mathrm{i}}(L)$ and the distribution of $L$~\cite{support}, including puddles of all sizes~\cite{Puddle}. We find  $<\varepsilon_{\mathrm{ij}}>$ = 110 K and 9 K for $V_{0}$ = 43 meV and 13 meV respectively. In real samples $V_{0}$ obeys a Gaussian distribution~\cite{support}. Statistically 25\% of all puddles have $V_{0} < $ 13 meV and 68\% of all puddles obey $V_{0} <$ 43 meV. Thus these calculations provide reasonable bounds for $<\varepsilon_{\mathrm{ij}}>$ in our samples. The above analysis suggests that $<\varepsilon_{\mathrm{ij}}>$ is likely to be a few tens of Kelvin, which is in reasonable agreement with our extracted $E_2\approx$ 10 K and supports the NNH scenario.

The above model points to the onset of VRH conduction at yet lower temperature. Indeed, $R_{\mathrm{NP}}(T)$ data in Fig.~\ref{RNPdata}(a) shows a bend at 5 K. We have verified that this bend is not due to current heating. Although the temperature range is limited, the VRH term in Eq.~\ref{fit} is necessary to capture this behavior. At large $|V^{\prime}_{\mathrm{bg}}|$, the extracted $T_{3}$ ranges 0.7 K -- 1.3 K as shown in Fig.~\ref{result}(c). These values are in very good agreement with $T_{3}$ = 0.5 -- 0.8 K obtained in Ref.~\onlinecite{Oostinga2008}, where the VRH conduction is observed in similar samples for a larger temperature range 5 K $> T >$ 50 mK. At small $|V^{\prime}_{\mathrm{bg}}|$, $T_{3}$ is so small that $R_{\mathrm{NP}}(T)$ remains essentially $T$-independent, probably due to Klein tunneling.

The corresponding $R_{3}$ is $\sim$8 k$\Omega/\square$ at small $|V^{\prime}_{\mathrm{bg}}|$ and increases with increasing $|V^{\prime}_{\mathrm{bg}}|$ in all samples. $R_{3}$ reaches approximately 75 k$\Omega/\square$ in samples A1 and A2 at $|V^{\prime}_{\mathrm{bg}}|$ = 90 V. This trend is consistent with the increasing suppression of Klein tunneling as $\Delta$ increases. The opening of the gap may also alter the strength of the electron-acoustic phonon coupling, which is essential to the VRH process, and affect $R_{3}$. As a final check, we note that VRH is expected to occur at $T << T_{\mathrm{c}}$, where $T_{\mathrm{c}}$ is determined by Eq.~\ref{2}, where the variable activation energy $E_{\mathrm{a}}$ becomes much smaller than $E_2$~\cite{Shklovskii1984}:
\begin{equation}
E_{a}(T_{\mathrm{c}})=\frac{\mathrm{d}(\mathrm{ln}R)}{\mathrm{d}(k_{\mathrm{B}}T)^{-1}}|_{T=T_{\mathrm{c}}}
=\frac{k_{\mathrm{B}}T_{3}^{\frac{1}{3}}T_{\mathrm{c}}^{\frac{2}{3}}}{3}\approx\frac{E_{2}}{3}
\label{2}.
\end{equation}
Using $T_{3}$ = 0.9 K and $E_{2}$ = 10 K, we obtain $T_{\mathrm{c}}$ = 33 K. Thus, the assignment of VRH at $T <$ 5 K is self-consistent.

In conclusion, we measure the $T$-dependent resistance at the charge neutrality point in gapped bilayer and develop a simple model to account for the observed multiple energy scales. Localized states produced by potential fluctuations play a crucial role in our model. At high temperatures, thermal activation to the mobility edge dominates the conduction. The activation energy is determined by the potential fluctuations at small bangap and approaches that of a band insulator at large bandgap or in clean samples. At lower temperatures, a percolation network forms via carrier hopping through localized states, leading to a second activated process and eventually variable range hopping. Our model provides important insights into electrical transport in gapped bilayer graphene, which may be useful in a range of electronic and optical applications.

\begin{acknowledgments}
We thank Brian LeRoy for sharing his STM data with us. We are grateful for a stimulating discussion with Pablo Jarillo-Herrero and thank Vin Crespi and Jainendra Jain for useful comments. This work is supported by NSF CAREER grant no. DMR-0748604 and NSF NIRT grant no. ECS-0609243. The authors acknowledge use of facilities at the PSU site of NSF NNIN.
\end{acknowledgments}


\begin{thebibliography}{30}
\bibitem[{R1()}]{R1}
\bibinfo{note}{The only ambiguous fitting cases are with the determination of
  $E_{1}$ and $R_{1}$ in samples A1 and A2 for $V^{\prime}_{\mathrm{bg}}$= -70,
  -80 and -90 V. There, $E_{1}$ and $R_{1}$ show strong correlations. We have
  fixed $R_{1}$ = 4,500 $\Omega$ based on results at smaller
  $V^{\prime}_{\mathrm{bg}}$ and determined $E_{1}$ accordingly.}
\bibitem[{Pud()}]{Puddle}
\bibinfo{note}{Puddles on the percolation path may have a narrower energy
  distribution and consequently lower $<\varepsilon_{\mathrm{ij}}>$. This
  effect has not been considered in our simulations.}
\end{thebibliography}

\end{document}